\documentclass[nofootinbib,pra,twocolumn,a4paper,groupedaddress,showpacs,floatfix,aps,longbibliography,10pt]{revtex4-1}
\usepackage{graphicx,amsmath,amssymb,amsfonts,dsfont,subfigure}

\newcommand{\ket}[1]{\ensuremath{|#1\rangle}}
\newcommand{\mc}[1]{\ensuremath{\mathcal{#1}}}
\newcommand{\bra}[1]{\ensuremath{\langle #1 |}}

\usepackage{color}

\newcommand{\vroW}{\varrho}

\begin{document}

\title{Lensing effect of electromagnetically induced transparency involving a Rydberg state}

\author{Jingshan Han,${}^1$ Thibault Vogt,${}^{1,2}$ Manukumara Manjappa,${}^1$ Ruixiang Guo,${}^1$  Martin Kiffner,${}^{1,3}$  and Wenhui Li${}^{1,4}$}

\affiliation{${}^1$Centre for Quantum Technologies, National University of Singapore, 3 Science Drive 2, Singapore 117543}
\affiliation{${}^2$MajuLab, CNRS-UNS-NUS-NTU International Joint Research Unit UMI 3654, Singapore, 117543}
\affiliation{${}^3$Clarendon Laboratory, University of Oxford, Parks Road, Oxford OX1 3PU, United Kingdom}
\affiliation{${}^4$Department of Physics, National University of Singapore, Singapore, 117542}



\begin{abstract}
We study the lensing effect experienced by a weak probe field under conditions of electromagnetically induced transparency (EIT) involving a Rydberg state. A Gaussian coupling beam   tightly focused on a laser-cooled atomic cloud produces an inhomogeneity in the coupling Rabi frequency along the transverse direction and makes the EIT area acting like a gradient-index medium. We image the probe beam at the position where it exits the atomic cloud, and observe that a red-detuned probe light is strongly focused with a greatly enhanced intensity whereas a blue-detuned one is de-focused with a reduced intensity. Our experimental results agree very well with the numerical solutions of Maxwell-Bloch equations.
\end{abstract}

\maketitle

%
%
\section{Introduction \label{introduction}}
The optical properties of a medium can be drastically modified by strong coherent interaction with a laser field, and one of the most prominent examples of the kind is electromagnetically induced transparency (EIT) \cite{fleischhauer2005electromagnetically}, which allows light transmission with large dispersion and gives rise to fascinating phenomena, such as extremely slow group velocity and light storage \cite{hau:99,kash:99,budker:99, chaneliere:05, eisaman:05}. Besides extensive investigations of the temporal dynamics, the spatial effects resulting from EIT have also been studied such as the focusing and de-focusing of transmitted probe light in the presence of a strongly focused coupling beam \cite{moseley1995spatial,moseley1996electromagnetically} and the deflection of probe light when passing through an EIT medium in the presence of a magnetic field gradient \cite{karpa2006stern,zhou2007deflection}. Recently, cancellation of optical diffraction was obtained for a specific detuning of the probe beam where the Doppler-Dicke effect compensates for diffraction \cite{firstenberg2009elimination,firstenberg2009eliminationB}.

While studies of EIT generally focus on $\Lambda$ type energy level configurations, more recently, there has been considerable interest with EIT in a ladder scheme involving Rydberg energy levels \cite{mohapatra2007coherent,pritchard2010cooperative, petrosyan2011electromagnetically} (Rygberg EIT). Strong dipolar interaction between Rydberg atoms in such EIT schemes is responsible for the so-called photon blockade, which offers promising means to realize deterministic single photon sources \cite{dudin2012strongly,peyronel2012quantum}, to induce effective interactions between photons \cite{firstenberg2013attractive}, and to realize photonic phase gates \cite{paredes2014all}. Rydberg EIT has also attracted attention with the demonstration of interaction enhanced absorption imaging (IEAI) \cite{gunter2012interaction,gunter2013observing}. This imaging technique detects Rydberg excitations via their modification on EIT transparency due to the strong interaction between Rydberg atoms. It confers great potential for the study of many-body physics with Rydberg atoms \cite{low2012experimental,weimer2010rydberg}.

Rydberg EIT experiments generally require strongly focused coupling fields in order to obtain sufficiently strong Rabi frequencies on the transition involving the Rydberg state. This focusing inevitably produces strongly inhomogeneous coupling fields. While lensing effect on the probe field associated with this inhomogeneity has been studied using a hot vapour \cite{moseley1995spatial,moseley1996electromagnetically}, until now this effect in cold Rydberg ensembles has received little attention. However, since the probe field is to be strongly modified by interaction induced nonlinearity in cold Rydberg ensembles, having a good understanding and control of the lensing effect is necessary.

We present in this paper a precise study of the lensing effect on the probe light by a tightly focused coupling beam in a Rydberg EIT scheme and its dependence on the probe detuning. In contrast to most previous studies on Rydberg EIT, the spatial structures are imaged in our experiment by a diffraction limited optical system. We use $27s$ Rydberg state of $^{87}$Rb atoms so that the effect of interaction between Rydberg atoms is minimal hence the experimental results can be accurately compared with numerical solutions of Maxwell-Bloch equations. This study sets clear delimitation on the possibilities offered by Rydberg EIT.

%
\section{Experiment \label{setup}}

\begin{figure}[hpb]
\includegraphics[width=8.1cm]{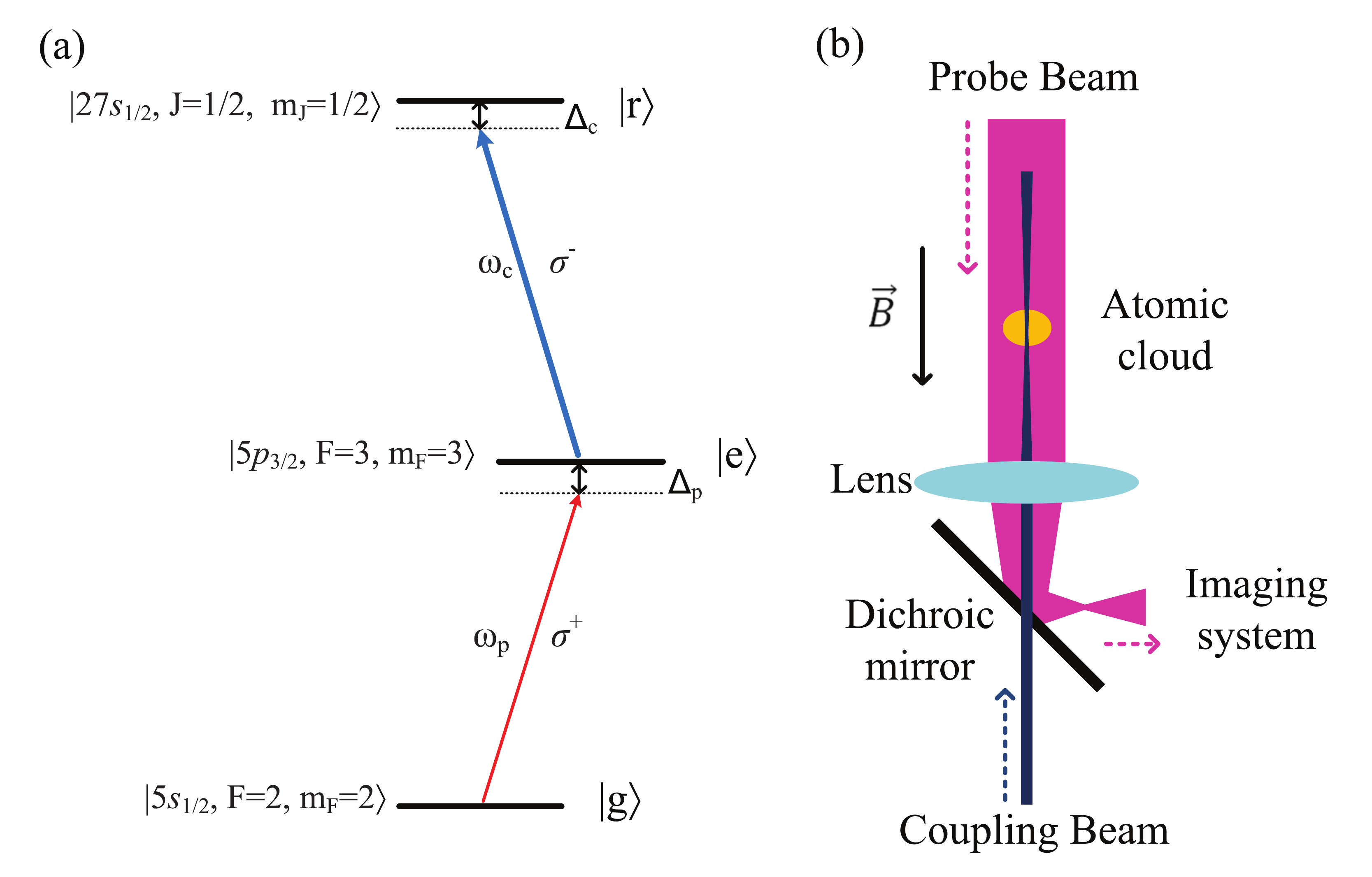}%
\caption{(a) The diagram of energy levels involved in the ladder scheme EIT. A probe light of $\sigma^+$ polarization drives the transition from $|5s_{1/2}, F=2, m_F=2\rangle$ ($|g\rangle$) to $|5p_{3/2}, F=3, m_F=3\rangle$ ($|e\rangle$), while a coupling light of $\sigma^-$ polarization drives the transition from $|5p_{3/2}, F=3, m_F=3\rangle$ to $|27s_{1/2}, m_J=1/2, m_I = 3/2\rangle$, which is not distinguishable in energy from other hyperfine states of $|27s_{1/2}, J=1/2, m_J=1/2\rangle$ ($|r\rangle$) in our setup. The detuning of the probe (coupling) light, $\Delta_p$ ($\Delta_c$) is defined as $\Delta_p = \omega_p - \omega_e$ ($\Delta_c = \omega_c - \omega_r$), where $\omega_p$ ($\omega_c$) is the frequency of the probe (coupling) light and $\omega_e$ ($\omega_r$) is the resonance frequency of the $\ket{e}\leftrightarrow\ket{g}$ ($\ket{r}\leftrightarrow\ket{e}$) transition. (b) The schematics of the optical setup for EIT beams. The magnetic field $\vec{B}$ along the vertical direction is pointing from top to bottom. The probe beam and the coupling beam are counter-propagating along the quantization axis, which is also the axial axis of the atomic cloud (indicated as a solid ellipse). After passing through the atomic cloud, the probe beam is separated from the coupling beam by a dichroic mirror and goes through the rest of optical imaging system to be imaged onto an electron multiplying charge coupled device (EMCCD camera). The lens shown here has a focal length of 160 mm. The dimensions are not to scale, but only indicate their relative shapes and positions. \label{ExperimentSetup}}
\end{figure}

The preparation of an ultracold $^{87}$Rb atomic sample for our experiment starts with loading a magneto-optical trap (MOT) from a Zeeman-slowed atomic beam, followed by further molasses cooling of the atomic cloud. Subsequently, a guiding magnetic field of approximately 3.5 Gauss along the vertical direction pointing downwards, as shown in Fig.~\ref{ExperimentSetup}(b), is switched on to define the quantization axis, and the atoms in the molasses are optically pumped into $|5s_{1/2}, F=2, m_F=2\rangle$ state for experiment. The population in $|5s_{1/2}, F=2, m_F=2\rangle$ state is controlled by de-pumping a certain fraction of atoms into $|5s_{1/2}, F=1\rangle$ level during this optical pumping stage. This de-pumping scheme allows varying the atomic density without changing much the atomic cloud size \cite{JDPritchardThesis}. At this stage, the atomic cloud has a temperature in the range of 28$\mu$K to 40$\mu$K.

A time of flight (TOF) of 6 ms following the optical pumping results in an atomic cloud that has a $1/e^2$ radius $w_r$=2.0 - 2.2 mm in the radial direction and a $1/e^2$ radius $w_z$=1.1 - 1.2 mm in the axial direction (along the quantization axis defined by the guiding B-field). The peak atomic density of $|5s_{1/2}, F=2,m_F=2\rangle$ state, $n_0$, can be varied from  $0.3 - 1.6 \times10^{10} \mathrm{cm^{-3}}$ .

\begin{figure}[hpb]
\includegraphics[width=8.1cm]{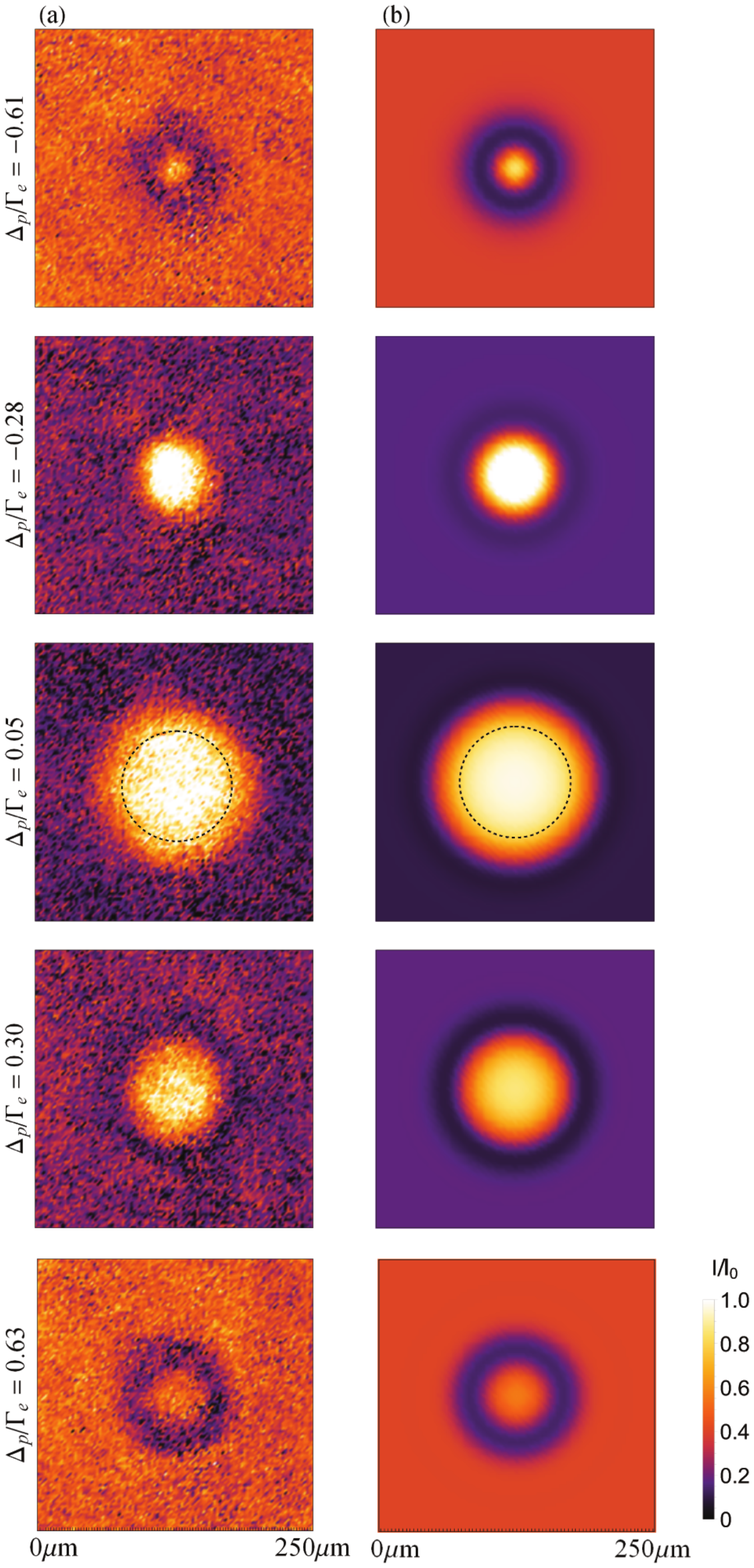}%
\caption{Images of the probe light from (a) experiment and (b) simulation. The probe detuning $\Delta_p$ for each set of images is given on the left side. The images in (a) are taken under the experimental conditions of $w_z$=1.1$\pm 0.1$ mm, $n_0$ = ($0.59\pm0.06)\times10^{10} \mathrm{cm^{-3}}$, $\Omega_{p0}/\Gamma_{e}=0.16\pm0.01$, $w_c$ = $49\pm1$ $\mathrm{\mu m}$, $\Delta_c/\Gamma_{e} = 0\pm0.05$, $\Omega_{c0}/\Gamma_{e}=1.98\pm0.05$. Each image in (a) is an average of 5 experimental shots. The same experimental conditions are also used as the inputs for solving the Maxwell-Bloch equations to generate the simulated images in (b), as detailed in the text. The thin dotted circles on the images of $\Delta_p/\Gamma_{e} = 0.05$ indicate the $1/e^2$ Gaussian size of the coupling beam. The color scale at the bottom right applies to all images. \label{Pic}}
\end{figure}

The states involved in the ladder scheme EIT are shown in Fig.~\ref{ExperimentSetup}(a), and the schematics of the optical setup for the EIT beams is shown in Fig.~\ref{ExperimentSetup}(b). The 780 nm laser beam for driving the $|g\rangle \rightarrow |e\rangle$ probe transition is generated from a Toptica DL pro diode laser, and the 480 nm laser beam for driving the $|e\rangle \rightarrow |r\rangle$ coupling transition is generated by a Toptica TA-SHG frequency-doubled diode laser system. Both the 780 nm laser and the 480 nm laser (via the fundamental light at 960 nm) are frequency locked to the same high-finesse Fabry-Perot cavity by Pound-Drever-Hall technique, which yields a linewidth of $\lesssim$ 30 kHz for the 780 nm laser and $\lesssim$ 60 kHz for the 480 nm laser. As illustrated in Fig.~\ref{ExperimentSetup}(b), the probe beam passing through the atomic cloud has a collimated $1/e^2$ radius $w_p$ of 3.45 mm, while the coupling beam is focused at the center of the atomic cloud with a $1/e^2$ radius $w_c$ in the range of 30 - 50 $\mu$m. When the incoming probe beam Rabi frequency $\Omega_{p0}$ is much smaller than the peak Rabi frequency of the coupling beam $\Omega_{c0}$, $\Omega_{p0} \ll \Omega_{c0}$, the coupling beam opens up a transparency window for the probe light to propagate through the otherwise opaque atomic cloud at the frequency around the probe transition resonance. It also induces a large index gradient along its transverse direction and results in a lensing effect. The intensity distribution of the probe beam at the exit of the atomic cloud, 1.1 mm below the center of the cloud, is directly imaged on the EMCCD camera through a diffraction limited optical system.

In each experimental cycle, the atomic cloud is prepared in $|5s_{1/2}, F=2,m_F=2\rangle$ state as described above, and the probe and coupling beams are turned on simultaneously for 15 $\mu$s during which the camera is exposed to take the image of the transmitted probe beam. To obtain an EIT transmission spectrum, the probe detuning $\Delta_p$ is varied from shot to shot to scan through the probe resonance while the coupling beam detuning $\Delta_c$ is fixed throughout. Shown in Fig.~\ref{Pic}(a) are a set of sample images of the transmitted probe light taken at different probe detunings $\Delta_p$, and the detailed description and discussion on the images and the spectra extracted from them are given in the next section.

\section{Results and discussion \label{results}}

\begin{figure}[hpb]
\includegraphics[width=8.1cm]{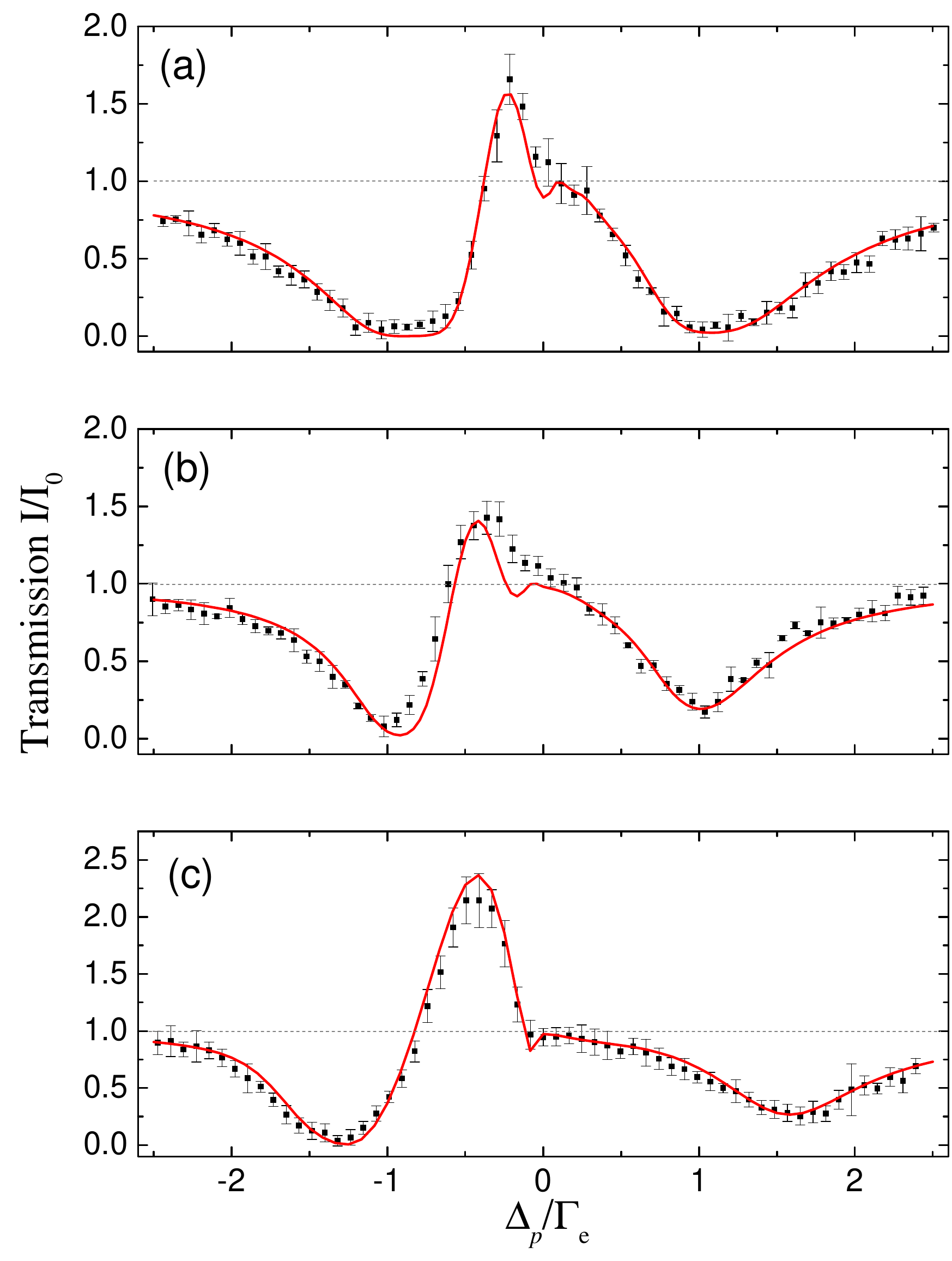}%
\caption{Transmission spectra of the transmitted probe light for different atomic densities and different coupling beam sizes. The black squares with error bar are experimental data, and the red lines are results of simulation that has only experimental parameters as input (please see the text). The spectra are taken at the conditions of (a) $w_z$=1.2$\pm 0.1$ mm, $n_0$ = ($1.40\pm0.15)\times10^{10} \mathrm{cm^{-3}}$,  $w_c$ = 49 $\pm$ 1 $\mathrm{\mu m}$, $\Delta_c/\Gamma_{e}$ = $0.16\pm0.05$, $\Omega_{c0}/\Gamma_{e} =1.98\pm0.05$; (b) $w_z$=1.1$\pm 0.1$ mm, $n_0$ = $(0.59\pm0.06)\times10^{10} \mathrm{cm^{-3}}$, $w_c$ = 49 $\pm$ 1 $\mathrm{\mu m}$, $\Delta_c/\Gamma_{e} = 0\pm0.05$, $\Omega_{c0}/\Gamma_{e} = 1.98\pm0.05$; (c) $w_z$=1.1$\pm 0.1$ mm, $n_0$ = $(0.69\pm0.07)\times10^{10} \mathrm{cm^{-3}}$, $w_c$ = 34 $\pm$ 1 $\mathrm{\mu m}$, $\Delta_c/\Gamma_{e} = 0\pm0.05$, $\Omega_{c0}/\Gamma_{e}=3.18\pm0.05$. All three spectra are taken with $\Omega_{p0}/\Gamma_{e}=0.16\pm0.01$.\label{spectra}}
\end{figure}

While different models have been developed to give accurate descriptions of the spatial effects of inhomogeneous EIT media on the propagation of the probe light \cite{moseley1996electromagnetically,manassah1996induced,zhou2007deflection,zhang2009birefringence}, the essential physics can be qualitatively captured in the following argument.

In EIT, the linear susceptibility for the probe light is given by
\begin{equation}	
\chi^{(1)} \left ( \vec r \right) =-i \frac{n_{at}\left ( \vec r \right) \Gamma_{e} \sigma_0 \lambda}{ 4 \pi \left(\gamma _{ge}-i \Delta _p+\frac{\Omega _c \left ( \vec r \right) ^2}{4 (\gamma _{gr}-i (\Delta _c+\Delta _p))}\right)} \label{susceptibility},
\end{equation}
where $\lambda$ is the wavelength of the probe transition, $\sigma_0=3 \lambda^2 /2 \pi$ the resonant cross-section of the probe transition,  $\Gamma_{e}=2\pi\times6.067$ MHz the decay rate of intermediate state $|e\rangle$, $\Delta _p$ and $\Delta _c$ the detunings of probe and coupling lights as defined earlier, and finally $\gamma _{ge}\approx\Gamma_e/2$ and $\gamma _{gr}=(\Gamma_r+\gamma _p+\gamma_c)/2+\gamma_D$ the decay rates of atomic coherences. Here, $\Gamma_r \sim 2\pi \times 10$ kHz \cite{branden2010radiative} is the decay rate of the upper state $|r\rangle$, $\gamma_{p} ( \gamma_{c} )$ is the linewidth of the probe (coupling) laser, and $\gamma_D$ is the dephasing rate from all other sources. The refractive index is related to the linear susceptibility $\chi^{(1)} \left ( \vec r \right)$ via the expression
  \begin{equation}
 n(\vec r) \approx \left( 1+\frac{1}{2} \Re \left(\chi^{(1)} \left( \vec r \right) \right) \right)\label{refractiveindex}.
\end{equation}

Seen from Eqs.\eqref{susceptibility} and \eqref{refractiveindex}, the inhomogeneity in atomic density $n_{at}\left( \vec r \right)$ and  Rabi frequency of coupling transition $\Omega_c \left( \vec r \right)$ can give rise to non-zero gradient in the refractive index, which results in the deflection of the probe light wave vector as it travels through such medium. For large $\Omega_c \left( \vec r \right)$, negligible $\gamma_{gr}$ and $\Delta_c\sim0$, the sign of the probe light detuning $\Delta_p$ decides the direction of the deflection either along or against the gradient of the refractive index.

In our experimental configuration, the atomic density $n_{at}(\vec r)$ along the radial direction of the transparency window is constant. On the other hand, the rapid change of the coupling Rabi frequency $\Omega_c(\vec r)$ due to the Gaussian intensity profile gives rise to a large gradient in the refractive index $n(\vec r)$. The probe light passing through this transparency window experiences lensing effects due to the high gradient of the refractive index, as can be seen in Fig.~\ref{Pic}.

The images in Fig.~\ref{Pic}(a) are acquired with the conditions detailed in the figure caption and from top to bottom, the probe detuning is varied from red to blue. The field of view of each image is centered around the coupling beam and is much smaller than the atomic cloud and the probe beam. The spot in the middle of each image is the transmitted probe light through EIT area while the uniform background indicates the absorption level of probe light by the atomic cloud with absence of the coupling light. It can be clearly seen that the intensity of the transmitted probe light at the red probe detuning $\Delta_p / \Gamma_e = -0.28$ is enhanced while the intensity on the blue side with a detuning $\Delta_p / \Gamma_e = 0.30$ is reduced, compared with the incoming probe intensity, which is about the same as the intensity of the transmitted probe beam on resonance ($\Delta_p / \Gamma_e = 0.05$ in Fig.~\ref{Pic}). Moveover, the spot size of the red-detuned probe light is smaller than that of the blue-detuned with a similar $|\Delta_p|$. Both the intensity and the size indicate the focusing of the red-detuned probe light and the defocusing of the blue-detuned one, since, if not due to the lensing effect, the transmitted spots would have similar intensity and size at detunings symmetric with respect to the resonance. It should be noted that the dark ring around the bright transmitted spots is not due to the lensing effect. Instead, it comes from the spatially varying coupling Rabi frequency as a result of the Gaussian intensity profile of the coupling beam. This spatial dependent Rabi frequency gives rise to a larger Autler-Townes splitting at the center of the coupling beam and smaller ones towards the edge of the beam. Consequently, the transmitted spot of on-resonance probe light ($\Delta_p / \Gamma_e = 0.05$ in Fig.~\ref{Pic}) has the largest size, since there is no Autler-Townes enhanced absorption throughout the whole EIT area, whereas the transmitted spots of off-resonance probe light have smaller sizes with surrounding dark rings due to enhanced absorption at Autler-Townes splitting frequencies. Because of this change of transmitted spot size vs. detuning, the focusing (defocusing) of the red (blue)- detuned probe light cannot be defined relative to the transmitted probe beam on resonance, but should rather be defined relative to the transmitted probe beam at that particular detuning with no lensing effect \footnote{The transmitted probe beam with no lensing effect can be simulated by removing the transverse gradient term of Eq.\eqref{maxS} in the appendix.}. In the experimental observation of Fig.~\ref{Pic} where the lensing effect is present, this can only be acknowledged by comparing the transmitted probe beam size and intensity at the detunings symmetric with respect to the resonance.

In order to obtain EIT transmission spectra, the transmission of the probe light is extracted by taking the ratio between the probe intensity at the center of such images ($I$) and that of the incoming probe beam without the atomic cloud ($I_0$). The transmission spectra shown in Fig.~\ref{spectra} are generated by plotting the probe transmission ($I/I_0$) as a function of the probe detunings $\Delta_p$ for atomic densities and coupling beam sizes detailed in the figure caption. As expected, there is a transparency spectral window near the probe resonance due to the coherent interaction between the coupling light and atoms, and the two absorption peaks are from the Autler-Townes splitting. The lensing effect within this transparency spectral range can be clearly seen from the greatly enhanced transmission at the red detuning $\Delta_p <$ 0 and the somewhat reduced transmission at the blue detuning $\Delta_p >$ 0. This enhanced transmission at the red detuning highly depends on the atomic density $n_{0}$ and the coupling beam size $w_c$. With the same coupling beam size $w_c$ and the same peak Rabi frequency $\Omega_{c0}$, the atomic cloud with a higher density in Fig.~\ref{spectra}(a) focuses the probe light more than that with a lower density in Fig.~\ref{spectra}(b). Moreover, if the atomic density is about the same, but the coupling beam size $w_c$ is focused down further (the peak Rabi frequency $\Omega_{c0}$ is consequently larger), the focusing of the probe light is greatly enhanced, as shown in Fig.~\ref{spectra}(b) and (c).

\begin{figure}[hpb]
\includegraphics[width=7.5cm]{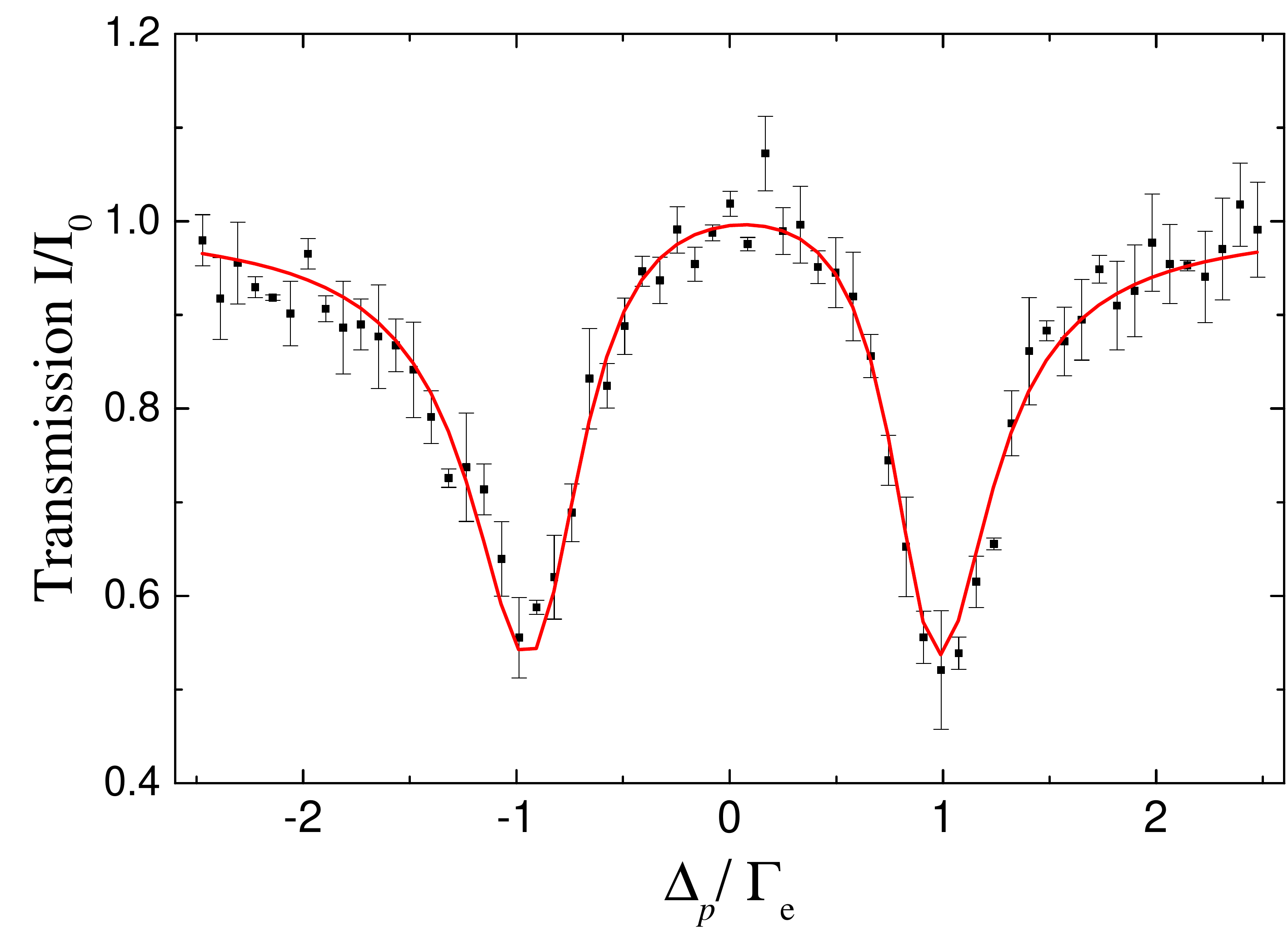}%
\caption{Transmission spectrum of the transmitted probe light through a thin single-beam ODT released atomic sample with the $1/e^2$ radius $w_z = 55.0\pm0.5$ $\mathrm{\mu m}$ (along propagation direction of EIT beams) and the atomic density $n_{0}$ = $(3.30\pm0.03)\times10^{10} \mathrm{cm^{-3}}$. The black squares with error bar are experimental data, and the red line is a one-dimensional fit from the formula $T(r=0,\Delta_p)$ given in the text.\label{ODTSpectra}}
\end{figure}

The lensing effect also critically depends on the size of the atomic cloud. In order to verify this, the same experiment is performed with an atomic cloud released from a very thin single-beam optical dipole trap (ODT). The ODT is horizontally positioned at 1.1 mm below the center of the molasses atomic cloud in the previous experiment, that is, at the object plane of the camera. The $1/e^2$ radius of this atomic cloud, which is along the propagation direction of the EIT beams, is $w_z=55.0\pm0.5$ $\mu$m (20 times smaller than that of the molasses atomic cloud). As shown in Fig.~\ref{ODTSpectra}, the spectrum of the Autler-Townes splitting is observed without any obvious lensing effect, in stark contrast to what is being observed in the molasses atomic cloud.

To understand our experimental results on lensing effect more quantitatively, we model our experimental system with a set of coupled Maxwell-Bloch equations as described in detail in the appendix. The inputs are the experimentally calibrated parameters including: a) the atomic density $n_{0}$ and the atomic cloud size $w_z$; b) the peak Rabi frequency $\Omega_{c0}$ and the waist $w_c$ of the coupling light; c) the initial Rabi frequency of the probe light $\Omega_{p0}$; d) the decay rate of atomic coherence $\gamma_{gr}$. The atomic density $n_{0}$ and atomic cloud size $w_z$ are well known from measurement with absorption imaging. The peak Rabi frequency $\Omega_{c0}$ and waist $w_c$ are extracted from a two-dimensional fit of images taken in the experiment performed on the thin atomic cloud released from the ODT. The transmission formula used for fitting is $T(r,\Delta_p)= \mathrm{exp}\left(-k\int_{-\infty}^{+\infty} \mathrm{Im}[\chi^{(1)}(r,z,\Delta_p)]dz\right)$, where $r$ and $z$ stand for the radial and axial coordinates respectively, $k=2\pi/\lambda$, and $\chi^{(1)}$ is defined in Eq.\eqref{susceptibility}. These parameters are given in the caption of Figs.~\ref{Pic} and~\ref{spectra}. The dacay rate of atomic coherence $\gamma_{gr}$ is obtained from fitting the probe beam transmission vs. $\Omega_{p0}$ at the center of the transmitted beam (with $\Omega_{c0}$) and $\Delta_p = 0$. This measurement yields the value of $\gamma_{gr}$ in the range of 50 - 150 kHz depending on atomic density, which is consistent with the evaluation of $\gamma_{gr}$ from various dephasing mechanisms in our experiment. $\gamma_{gr}$ used in the simulation is set to be 100 kHz since we find the results of the simulation are not very sensitive to $\gamma_{gr}$ in the range of 50 - 150 kHz. The simulated images of the probe light intensity at the exit of the atomic cloud are shown in Fig.~\ref{Pic}(b), along the side of the experimental images taken with the same parameters. The spectra from simulation are plotted together with experimental data in Fig.~\ref{spectra}. The experimental and theoretical results show excellent agreement, which confirms the good control in our experiment and lays a solid foundation for further pursuing the experimental investigation of the interaction between Rydberg excitations using IEAI in our system.

\section{Summary \label{summary}}
In summary, we have observed the lensing effect on the probe light in electromagnetically induced transparency involving a Rydberg state by directly imaging the probe beam passing through a laser-cooled atomic cloud. With the atomic cloud of only moderate optical depth, the transmitted probe light is strongly focused at a frequency red detuned from the probe resonance, and has a peak intensity a few times that of the input probe light. This study is important for imaging Rydberg excitations via interaction enhanced absorption imaging based on Rydberg EIT. It is also highly relevant in studying non-linearity of cold interacting Rydberg ensembles as the probe intensity determines the strength of interaction between Rydberg polaritons \cite{firstenberg2013attractive,bienias2014scattering}.  It will be interesting to investigate how such lensing effect is modified by the interaction between Rydberg atoms, which will be significant when a Rydberg state of high principal quantum number is used. Combining dispersive non-linearities and focusing, one may imagine creating a one-dimensional gas of Rydberg atoms. It may be also possible to tune the interaction between Rydberg atoms in order to switch from focusing to defocusing lensing effect.
%
\begin{acknowledgments}
The authors thank Thi Ha Kyaw, Nitish Chandra, and Armin Kekic for the early preparation of experimental setups, and acknowledge the support from the Ministry of Education and the National Research Foundation, Singapore. This work is partly supported through the Academic Research Fund, Project No. MOE2015-T2-1-085.
\end{acknowledgments}

\appendix
\section{Theoretical Model \label{Tmodel}}
We describe  the interaction of the probe and coupling fields with an
ensemble of ultracold atoms using the standard framework of coupled Maxwell-Bloch
equations. In electric-dipole and  rotating-wave approximation, the
Hamiltonian of each atom interacting with the probe and coupling fields is
\begin{align}
H  = & - \hbar\left(\Delta_p \ket{e}\bra{e}+ (\Delta_c+\Delta_p)\ket{r}\bra{r}\right) \notag \\
& -\hbar \left( \frac{\Omega_p}{2} S_e^+ + \frac{\Omega_c}{2} S_r^+ \,+\,\text{H.c.}\right)\, ,
\label{H}
\end{align}
where $\Delta_p$ ($\Delta_c$) is the probe (coupling) field detuning,
\begin{align}
\Delta_p=& \omega_p -\omega_{e}\,, \\
\Delta_c=& \omega_c -\omega_{r}\,,
\end{align}
and $\omega_{e}$ ($\omega_{r}$) is the resonance frequency on the $\ket{e}\leftrightarrow\ket{g}$
($\ket{r}\leftrightarrow\ket{e}$) transition. The atomic transition operators
$S_{x}^+$ ($x\in\{r,g\}$) in Eq.~(\ref{H}) are defined as
\begin{align}
S_e^+= \ket{e}\bra{g}, \quad S_r^+= \ket{r}\bra{e}.
\end{align}
The probe field Rabi frequency $\Omega_p$ inside the medium is a dynamical variable
that we want to determine at each position in space. On the contrary, the coupling field is almost unaffected by the medium, hence we assume that the spatial variation of  the coupling field is
\begin{align}
 \Omega_c =i\frac{ \Omega_{c0}  z_0}{z+i z_0} e^{-i  z_0 r^2/[w_{c}^2(z+iz_0)]} \,,
 \label{gaussian}
\end{align}
where $z_0$ is the Rayleigh length and $w_c$ is the beam waist at $z=0$.
The field in Eq.~(\ref{gaussian}) is a solution of Maxwell's equations in paraxial approximation and
in free space.
We model the time evolution of the atomic density operator $\vroW$
by a Markovian  master equation~\cite{breuer:os},
\begin{align}
\partial_t \vroW &= - \frac{i}{\hbar} [ H , \vroW ] +\mc{L}_{\gamma}\vroW+\mc{L}_{D}\vroW\,.
\label{master_eq}
\end{align}
The term $\mc{L}_{\gamma}\vroW$ in Eq.~(\ref{master_eq}) accounts for spontaneous
emission of the excited states. These processes are described
by standard Lindblad decay terms,
\begin{align}
\mc{L}_{\gamma}\vroW = &          -  \frac{\Gamma_e}{2}
\left( S_e^+S_e^-  \vroW  + \vroW S_e^+S_e^-
- 2 S_e^-\vroW S_e^+ \right) \notag  \\
& -  \frac{\Gamma_r}{2}
\left( S_r^+S_r^- \vroW  + \vroW S_r^+S_r^-
- 2 S_r^- \vroW S_r^+ \right) \, ,
\end{align}
where  $S_{x}^- = \left(S_{x}^+\right)^{\dagger}$ ($x\in\{r,g\}$) and $\Gamma_e$
is the full decay rate of state $\ket{e}$.  The long-lived Rydberg state $\ket{r}$
decays with $\Gamma_r\ll\Gamma_e$.
The last term  $\mc{L}_{D}\vroW$ in Eq.~(\ref{master_eq}) describes decoherence due to laser noise and is given by
\begin{align}
\mc{L}_{D}\vroW = & -  \frac{\gamma_c}{2}
\left( S_r^+S_r^- \vroW  + \vroW S_r^+S_r^-
- 2 S_r^+S_r^- \vroW S_r^+S_r^- \right)   \notag \\
& -  \frac{\gamma_p}{2}
\left( S_e^-S_e^+ \vroW  + \vroW S_e^-S_e^+
- 2 S_e^-S_e^+ \vroW S_e^-S_e^+ \right)   \,,
\end{align}
where $\gamma_c$ ($\gamma_p$) is the linewidth associated with the control (probe) field.

In  paraxial approximation and for a probe field  varying slowly in time Maxwell's equations reduce to
\begin{align}
& \left(-i\frac{c}{2 \omega_e}\Delta_{\perp}+ \partial_{z}+\frac{1}{c} \partial_{t}\right)  \Omega_{p}
=  i \eta \vroW_{eg}\,,
\label{maxS}
 \end{align}
where $\Delta_{\perp}$ is the transverse Laplace operator, $c$ is the  speed of light and the
coupling constant $\eta$ is defined as
\begin{align}
 \eta & =\frac{n_{at} \sigma_0}{2}\Gamma_e \,.
 \end{align}

The set of equations~(\ref{master_eq}) and~(\ref{maxS}) represent a system of
coupled, partial differential equations and have to be solved consistently for
given initial and boundary conditions.
Here we consider the steady state regime and find  the time-independent solution $\vroW^{\text{st}}$
to Eq.~(\ref{master_eq}). Note that $\vroW^{\text{st}}$ solves Eq.~(\ref{master_eq}) to all orders in the
 probe field, and $ -2 \eta \vroW^{\text{st}}_{eg}/ k \Omega_p$ reduces to the linear susceptibility given
 in Eq.~(\ref{susceptibility}) only for  a weak probe field $\Omega_p\ll\Gamma_e$.
We replace $\vroW_{eg}$ in Eq.~(\ref{maxS}) by the non-perturbative
expression for $\vroW^{\text{st}}_{eg}$ such that Eq.~(\ref{maxS}) reduces to a nonlinear, time-independent
equation for the probe field Rabi frequency.
We numerically solve this equation in three spatial dimensions with the software
packet MATHEMATICA~\cite{MM}  and the implicit
differential-algebraic solver (IDA) method option for NDSolve.
%
%

%

%
%
\end{document}